\documentclass[fleqn,12pt,twoside]{article}
\usepackage{espcrc1}

\usepackage{graphicx}
\usepackage[figuresright]{rotating}

\newcommand{\AmS}{{\protect\the\textfont2
  A\kern-.1667em\lower.5ex\hbox{M}\kern-.125emS}}

\title{Dynamics of Multiquark Systems: Mass, Width and Exotics}

\author{Makoto Oka\address{Department of Physics, H-27, Tokyo Institute of Technology\\
Meguro, Tokyo 152-8551 Japan}\thanks{email: oka@th.phys.titech.ac.jp}}
       
\begin{document}

\maketitle

\begin{abstract}
Exotic multi-quark states are examined in the quark model and in QCD.
Current status of theoretical studies of the pentaquark $\Theta^+$ is reported.
We show recent analyses of multi-quark components of baryons.
A novel method to extract a compact excited state in the lattice QCD is applied.
We also discuss how to determine mixings of different Fock components in the hadron
in the QCD sum rule approach.
\end{abstract}

\section{Introduction: Quark Model}

The quark model describes mesons and baryons in terms of constituent quarks.
Mesons are $q\bar q$ states and baryons are $qqq$, while all the other combinations of
quarks, such as $qq\bar q\bar q$, $qqqq\bar q$, are called exotic.
The constituent quarks must be effective degrees of freedom which are valid only 
in low energy regime. 
They must have the same conserved charges as the QCD quarks: 
baryon number 1/3, spin 1/2, color 3 and flavor 3.
They acquire dynamical masses of order $\sim 300-500$ MeV
induced by chiral symmetry breaking of the QCD vacuum.

The ground-state mesons and baryons have been classified 
based on $SU(3)\times SU(2) \to SU(6)$ symmetry, i.e., 
the pseudoscalar and vector meson nonets, and
the octet and decuplet baryons.
Their mass spectra, in particular, the pattern of SU(3) breaking,
and the electro-magnetic properties
are well reproduced by minimum dynamics of quarks,
quark confinement and color-magnetic type interactions,
except for pseudoscalar mesons, i.e., the octet pseudoscalars, $\pi, K, \eta$,
are very light according to chiral symmetry breaking,
while the $\eta'$ mass is large due to the $U_A(1)$ anomaly.

The quark model based on $SU(6) \times O(3)$ symmetry, however, encounters some difficulties when it is applied to
excited hadron states.
An example is the scalar meson nonet; ($\sigma(600), a_0(980), f_0(980), K_0^*(800)$).
($K_0^*$ has been indicated in $K\pi$ final states in $J/\psi$ and $D$ meson decays, 
but not yet established\cite{kappa}.)
Their mass ordering as $q\bar q$ states is expected to be like
$m(\sigma) \sim m(a_0) < m(f_0)$, assuming the ideal mixing, i.e.,
\begin{eqnarray}
&& \sigma\sim \frac{u\bar u + d\bar d}{\sqrt{2}} , \qquad
a_0\sim \frac{u\bar u - d\bar d}{\sqrt{2}} , \qquad
f_0 \sim s\bar s \nonumber
\end{eqnarray}
Furthermore, while they are classified as $^3P_0$ states, 
their spin-orbit partners $J= 1$ and 2 states are not observed in their vicinity.

A possible solution of the difficulty is to consider four-quark exotic states for the scalar mesons\cite{4q}.
Suppose that diquarks with flavor 3, color 3 and spin 0, i.e.,
\begin{eqnarray}
&& U=(\bar d\bar s)_{S=0,C=3,f=3}\qquad D=(\bar s\bar u)_{S=0,C=3,f=3}
\qquad S=(\bar u\bar d)_{S=0,C=3,f=3} . 
\label{diquark}
\end{eqnarray}
are building blocks of the scalar mesons.
Then the scalar nonets in the ideal mixing may appear as
\begin{eqnarray}
&& \sigma\sim S\bar S \sim (ud)(\bar u\bar d) \nonumber\\
&& a_0\sim \frac{1}{\sqrt{2}} (U\bar U - D\bar D) \sim \frac{1}{\sqrt{2}} ((ds)(\bar d \bar s)-(su)(\bar s\bar u))
\nonumber\\
&& f_0 \sim \frac{1}{\sqrt{2}} (U\bar U + D\bar D) \sim \frac{1}{\sqrt{2}} ((ds)(\bar d \bar s)+(su)(\bar s\bar u))
\nonumber
\end{eqnarray}
Then one sees that the strange quark counting predicts the observed mass pattern,
$m(\sigma) < m(a_0) \sim m(f_0)$.
It also explains that the $J=0$ state is isolated without $J=1, 2$ partners.

Thus one sees that multi-quark components may help to explain anomalies in the scalar meson 
nonets.  There are other hadrons which are suspected to contain exotic multi-quark components, such as 
$D_s^*$, $X(3872)$ and $\Lambda(1405)$, 
mainly because they do not fit well in the spectra of the ordinary mesons or baryons.

The next question is whether the QCD dynamics allows such states?
We actually have a simple reason,  for instance, why $\Lambda (1405)$ 
is possible to be a 5-quark state.
$\Lambda(1405)$ is a $J^{\pi} = 1/2^-$ flavor singlet baryon.
The three quark ($uds$) configuration has to contain orbital excitation $L=1$ with spin 1/2,
leading to $J=1/2^-$ and $3/2^-$ states. 
Then the candidate of the $3/2^-$ partner is $\Lambda(1520)$, but the spin-orbit splitting is
unusually large compared to the nonstrange baryons with $L=1$.
On the other hand, the 5-quark content, $udsu\bar u+udsd\bar d$ may be 
realized in $L=0$ without orbital excitation,
and thus has advantage over the $L=1$ excited states.
In terms of the diquark language, one may assign the $L=0$ and $S=1/2$ configuration.
This gives an isolated $J=1/2$ state and has no difficulty of large LS splitting.
Therefore it is quite interesting and important to answer whether realistic dynamical calculation
indeed gives multi-quark states a lower energy.

\section{Pentaquark $\Theta^+$}
We first look at the status of the pentaquark $\Theta^+$ in the quark picture.
The pentaquark $\Theta^+$ is a baryon with strangeness $+1$, whose minimal quark content
is $uudd\bar s$.  It may be affiliated to a member of flavor antidecuplet\cite{Theta-quark}.
How does the known dynamics of the quark model work for $\Theta^+$?
It has been clarified that the ``standard'' quark model, which explains the ground state mesons and baryons very well, does not easily yield the small mass and the tiny width of $\Theta^+$ simultaneously.
Recently variational techniques for solving multi-body bound states have been applied to the 5-quark systems. The results\cite{Theta-variational} show that 
the typical masses of spin $1/2^{\pm}$ and $3/2^{\pm}$ pentaquarks are
more than 500 MeV above the threshold of $NK$, and thus the masses are 1.9 GeV or higher.
Furthermore, if they have such large masses, various decay channels are open and their widths
must be very large.  It was pointed out that even at the mass 1.54 GeV, the decay widths of
$1/2^-$ 5-quark states\cite{Hosaka-width}, 
which decay into S wave $NK$  states, must be a few hundred MeV or larger.
 
In order to bring such high-mass resonances down to the observed mass\cite{Nakano}, 
one may require a very strong correlation. 
Diquark correlation is a possibility\cite{Theta-diquark}. 
Most dynamical models of quarks, such as one-gluon exchange, instanton induced interaction, and so on,
give attraction to color $\bar 3$, flavor $\bar 3$, spin 0 diquarks (Eq.(\ref{diquark})).
It is, however, noted that the same interaction causes (often stronger) attraction to $q \bar q$ color singlet, flavor 8, spin 0 system (i.e., pseudoscalar mesons). Then the ground state of the pentaquark may well be 
a state of a baryon and a meson far separated.
Thus, the di-quark scenario may not work unless the spin (or other quantum numbers) is chosen so that
it hinders a pseudoscalar subsystem. One such case is that the spin of $\Theta^+$ is 
3/2\cite{Hosaka-width,spin3_2}.

Even if the observed $\Theta^+$ might not be what we expected originally,
the techniques developed in the study of the pentaquark $\Theta^+$ are useful
in studying other exotic multiquark resonances as well as multiquark components of ordinary hadrons.
In particular, the variational method is powerful and useful to judge 
whether a certain quark model allows multiquark hadrons as a ground state.

\section{Pentaquarks in Lattice QCD}

As the model calculations have various ambiguity in treating multiquark systems with color singlet sub-systems,
direct applications of QCD to exotics are desperately needed.
In fact, QCD does not a priori exclude exotic multi-quark bound (resonance) states as far as they are color-singlet.
We here show some of the results from the lattice QCD simulations and the approach using QCD sum rules.

Lattice QCD is powerful in understanding non-perturbative physics of QCD: 
vacuum structure, phase diagrams, mass spectra of ground state mesons and baryons, interactions of quarks, . . . 
However, at this moment the lattice QCD has two rather severe restrictions:
(1) Light quarks are too expensive. The simulations are made at a large quark mass and require (often drastic) extrapolation to physical quark masses for $u$ and $d$ quarks.
(2) No direct access to resonance poles is possible. 
It is hard to distinguish resonances from hadron scattering states. Real number simulations can not access complex poles.
Thus applications to exotic hadrons are yet limited.

We  have developed a new method, called hybrid boundary condition (HBC) method, to extract the 5-quark resonances out of the meson-baryon background\cite{Ishii}.
The basic idea of HBC is to apply anti-periodic boundary conditions to certain quarks,
which effectively raise the energy of the lowest-energy meson-baryon scattering state.
Comparing the results with the standard boundary condition with the hybrid one, one can determine
whether a state seen in the lattice simulation is a compact resonance state or a hadronic continuum state.
This technique has been applied to the pentaquark $\Theta^+$ first and later to 
$\Lambda(1405)$ and the other possible exotics.

The results for $\Theta^+$ pentaquark are summarized as follows.\\
(1) The negative parity $1/2^-$ state appears at $m\sim 1.75$ GeV,
which seems consistent with $NK$ ($L=0$) scattering state on the lattice.
The HBC analysis confirms that the observed state is not a compact 5-quark state.\\
(2) The positive parity $1/2^+$ state appears at 2.25 GeV, which is too heavy for $\Theta^+$.\\
(3) The $J =3/2$ states are generated by three different operators: di-quark type, 
$NK^*$, and color-twisted $NK^* $ operators.
The results show that the mass of $3/2^-$ state is around 2.11 GeV, 
that is consistent with $NK^*$ ($L=0$) threshold, while the positive parity $3/2^+$ is around 2.42 GeV, consistent with $NK^*$ ($L=1$) threshold.

Thus no candidate for compact 5-quark state is found. 
Most lattice QCD parties agree with these results with a few exceptions.
The QCD sum rules also give the consistent results.

A similar method has been applied to study the penta-quark nature of
the flavor singlet $\Lambda (1/2^-)$ state\cite{LQCD-Lambda}.  
In this case, one may twist the spatial boundary condition 
only of quarks in the quenched approximation,
while antiquarks satisfy the standard boundary condition.
This ``HBC'' will enable us to isolate $\Lambda^*$ resonance 
from $N\bar K$ and $\Sigma\pi$ scattering states. 
The results are satisfactory, showing that the 5-quark operator 
gives $m_{5Q}$ = 1.63(7) GeV, which is around the $m(N)+m(K)$ threshold 
on the current choice of the lattice parameters.
It is further interesting that the HBC analysis indicates that this 5-quark state 
seems a compact resonance and thus it is a strong candidate for $\Lambda(1405)$.
In contrast, the 3-quark operator gives a higher mass, $m_{3Q}$ = 1.79(8) GeV.

\section{Exotic Multi-quark States in QCD Sum Rule}

In nature, it is expected that exotic multi-quark components mix with the standard Fock
state in hadrons. In such mixed states, one may ask how large the mixing probability is
of the exotic multi-quark components.

It turns out that such a mixing is not easily quantified.
It would be natural to consider the strengths of the couplings of 
the 3-quark and 5-quark 
operators to the physical state and then evaluate the mixing angle.
However, such a procedure is largely dependent on the definition
and normalization of the local operators.  Indeed, a numerical factor
can be easily hidden in the local operators and thus the magnitudes
of the coupling strengths are ambiguous.

This problem happens to be more fundamental than just 
the definition of the operators. 
In the field theory one may not be able to ``measure'' the number of quarks without
ambiguity because no conserved charge corresponding the number
of quarks, $N(q)+N(\bar q)$, is available.
Namely, the Fock space separation may not be unique.
Thus we have to consider the concept of the ``number of quarks''
in the context of the quantum mechanical interpretation of the 
field theoretical state.

In recent study\cite{SINNO}, we propose two ways to ``define'' the ratio of the Fock space
probability.
In the first approach, we define local operators
in the context of a 5-quark operator $J_5$.
As the 5-quark operator contains $qqq$ component, 
one can write the operator into $J_5= \tilde J_5 + \tilde J_3$.
Then the mixing parameter may be defined as the ratio of the couplings 
to $\tilde J_5$ and $\tilde J_3$.
This mixing angle can be evaluated from the correlation functions
with an assumption that the poles are at the same position.
The result is model independent, but it depends on the choice of the
operators.
Therefore it does not necessarily have a direct relation to the mixing parameters
employed in the quark models.

In order to define a mixing angle more appropriate to the quark models, 
one must determine the normalization of the operators using a quark model
wave function.
One may use the MIT bag model wave functions for the normalization,
assuming that the bag model states (with define number of quarks) are normalized
properly.

The above two methods have been applied to the $a_0$ scalar meson and we have
found that the physical state for $a_0$ is indeed given by the mixings of 
$q\bar q$ and $qq\bar q\bar q$ states.
It turns out that the 4-quark mixing probability is about 90\%, which does not 
strongly depend on the choice of the definition.

In contrast, for the flavor singlet $\Lambda^* (1/2^-)$ state, we find that the 3-quark operator gives 
a higher mass and thus the lowest energy state is predominantly 5-quark state.  In such a case,
the mixing is not properly defined in our method.

\section{Conclusion}

A technique using the hybrid boundary conditions seems to work in lattice QCD to distinguish compact states from scattering states.
The quenched lattice QCD suggests that $\Lambda(1405)$ is predominantly a 5-quark state.
The mixing of the multi-quark components is not quantified from the field theoretical viewpoint.
However, one can define a set of useful mixing amplitudes using the well-defined matrix elements of local operators. Whether this definition of the mixing is relevant in the quark model is yet an open problem. 
The QCD sum rule indicates a large 4-quark components in scalar mesons. 

\bigskip
The contents of this paper come from various collaborations.
I acknowledge Drs. N. Ishii, H. Suganuma, T. Doi, Y. Nemoto, H. Iida, F. Okiharu, S. Takeuchi, A. Hosaka,
T. Shinozaki, T. Nishikawa, J. Sugiyama,  and T. Nakamura for productive collaborations and discussions.


\begin{thebibliography}{9}
\bibitem{kappa} Particle Data Group, Jour. of Phys. {\bf G33} (2006) 1.
\bibitem{4q} R.L. Jaffe, Phys. Rev. {\bf D15} (1977) 267.

\bibitem{Theta-quark} M. Oka, Prog. Theor. Phys. {\bf 112} (2004) 1; R.L. Jaffe, Phys. Rept. {\bf 409} (2005)1.
\bibitem{Theta-variational} S. Takeuchi, K. Shimizu, Phys. Rev. {\bf C71} (2005) 062202;
E. Hiyama et al., Phys. Lett. {\bf B633} (2006) 237.
\bibitem{Hosaka-width} A. Hosaka, M. Oka, T. Shinozaki, Phys. Rev. {\bf D71} (2005) 074021.
\bibitem{Nakano} T. Nakano et al., LEPS collaboration, Phys. Rev. Lett. {\bf 91} (2003) 012002.
\bibitem{Theta-diquark} R.L. Jaffe, F. Wilczek, Phys. Rev. Lett. {\bf 91} (2003) 232003.
\bibitem{spin3_2} S. Takeuchi, K. Shimizu, Phys. Rev. {\bf C71} (2005) 062202.
\bibitem{Ishii}  N. Ishii et al., Phys. Rev. {\bf D71} (2005) 034001; N. Ishii et al., Phys. Rev. {\bf D72} (2005) 074503.
\bibitem{LQCD-Lambda} Y. Nemoto et al., Phys. Rev. {\bf D68} (2003) 094505; N. Ishii et al., in preparation.
\bibitem{SINNO} J. Sugiyama et al., in preparation.

\end{thebibliography}
\end{document}